# Quantitative Assessment of Adulteration and Reuse of Coconut Oil Using Transmittance Multispectral Imaging

S. Herath[a], H.K. Weerasooriya[a], D.Y.L.Ranasinghe[b], W.G.C.Bandara[a], H.M.V.R.Herath[a], G.M.R.I.Godaliyadda[a], M.P.B.Ekanayake[a], Terrence Madhujith[c]

[a] Department of Electrical and Electronic Engineering, Faculty of Engineering, University of Peradeniya, Peradeniya, Sri Lanka (20400)
[b] Office of Research and Innovation Services, Sri Lanka Technological Campus, Padukka, Sri Lanka (10500)
[c] Department of Food Science and Technology, Faculty of Agriculture, University of Peradeniya, Peradeniya, Sri Lanka (20400)

*Abstract*— **Coconut oil known for its wide range of uses is often adulterated with other edible oils. Repeated use of coconut oil in food preparation could lead to many health issues. Existing methods available for evaluating quality of oil are laborious and time consuming. Therefore, we propose an imaging system hardware and image processing-based algorithm to estimate the adulteration of coconut oil with palm oil as the adulterant. A clear functional relationship between adulteration level and Bhattacharyya distance was observed as $R^2 = 0.9876$ on the training samples. Thereafter, another algorithm is proposed to develop a spectral-clustering based classifier to determine the effect of reheat and reuse of coconut oil. Distinct clusters were obtained for different levels of reheated oil classes and the classification was performed with an accuracy of 0.983 on training samples. Further, the input images for the proposed algorithms were generated using an in-house developed transmittance based multispectral imaging system.**

***Keywords*— Adulteration; Coconut oil quality; Food quality estimation; Multispectral imaging; Repeatedly used coconut oil.**

## I. INTRODUCTION

Coconut (*Cocos nucifera*), belonging to the palm family is a multipurpose tree with many uses. The fibrous one-seeded drupe is used for the production of coconut water, coconut milk, desiccated coconut, and coconut oil. Coconut oil has been used as a cooking or frying oil, as an ingredient in some foods, production of skincare products, pharmaceuticals, among others. Palm oil which shows quite similar physical properties is often used to adulterate coconut oil (Young, 1983;Pandiselvam et al., 2019 ) as their cost of production is significantly less than that of coconut oil. Furthermore, the use of frying oil over and over many a time is common in foodservice establishments and at the domestic level to cut down on the cost. However, unfortunately, the chemical and thermophysical properties are altered during reuse (Bhuiyan et al., 2016) and these physico-chemical changes compromise the safety of edible oils and, thus making fried foods unsafe for consumption (De Alzaa F et al., 2018). During repeated heating, many secondary oxidative products such as carbonyls, organic acids, hydrocarbons and polymerized compounds and thermolyzed constituents are generated. Furthermore, *trans* fatty acids which raise LDL and total cholesterol while decreasing HDL cholesterol are also generated. As a result, increased risk of cardio vascular diseases, inflammation, oxidative stress, carcinogenesis and other non-communicable diseases (Hamsi et al., 2015) are possible due to long term use of adulterated and or reused coconut oil.

Chemical methods that are used to assess the quality of coconut oil are laborious, relatively expensive, time-consuming, and generate chemical waste, while the equipment used in spectroscopy methods, such as Raman and FTIR are expensive, thus involve high initial investment. ATR-FTIR spectroscopy was utilized in (Amit et al., 2020) for the rapid detection of coconut oil adulterated with fried coconut oil and in (Jamwal et al., 2019) for detection of adulteration of virgin coconut oil with paraffin oil. A method of assessing the quality parameters of edible oils such as authenticity, adulteration has been proposed in (Nunes, 2014) using the techniques vibrational spectroscopy and chemometrics. The capability of using the FTIR spectroscopy to determine the level of adulteration of virgin coconut oil by palm kernel oil was reported by (Manaf et al., 2007). To identify the authenticity of virgin coconut oil, the use of FT-MIR spectroscopy combined with chemometrics techniques were illustrated in a study (Rohman and Che Man, 2011).

The quality of food consumed plays a pivotal role in assuring the health of a society. Contamination and adulteration of food impose a serious threat to the quality of food. Therefore, it is of paramount importance to monitor the quality of food continuously. In this context, the need for accurate, fast, nondestructive, and economical methods to assure the quality of the food products is a timely need. The rapid development of computer-vision based techniques such as multispectral imaging (MSI) has resulted in such techniques being explored as viable options for various practical applications (Brosnan and Sun, 2004). Multispectral imaging is a rapidly developing, emerging scientific tool, which integrates both imaging and spectroscopic techniques into one. Minimal sample preparation, nondestructive nature and, fast acquisition times, are the main advantages of the MSI system over the traditional methods (Gowen et al., 2007). Spectral Imaging techniques have widely been used for the quality assessment of fruits and vegetables (ElMasry et al., 2008), snacks (Ramos-Diaz, 2019), fish (Cheng and Sun, 2015), and meat (Achata, 2020; Ma et al., 2019) VideometerLab spectral imaging instrument comprising of 18 bands has been used, to quantify the mold growth on food (Ebrahimi et al., 2015) and to detect the adulteration of pork and beef in raw meat (Ropodi et al., 2015). An MSI system operating in the spectral range of 405-970 nm consisting of 19 different spectral bands has been used to determine the aerobic plate count of cooked pork sausages in (Ma et al., 2014) . A hyperspectral microscopic imaging system with 300 spectral bands ranging from 400 – 998 nm was reported in (Ouyang et al., 2020), for assessment of matcha quality . A methodology to identify the level of adulteration in turmeric samples by examining the reflectance spectrum of multispectral images was presented by this research group previously (Chaminda et al., 2020). A prototype MSI system was developed to detect bruises on apples online(Huang et al., 2015). For most of the above studies, images were captured using complex hardware setups comprising of a high number of spectral bands and can acquire high-resolution images. Furthermore, algorithms used in the studies are based on Principal Component Analysis (Qu et al., 2015) for dimension reduction, K-means clustering (Ebrahimi et al., 2015) for clustering, Linear Discriminant Analysis (Femenias et al., 2020) and Support Vector Machine (Kamruzzaman et al., 2016) for classification and feature extraction.

The objective of the present study was to 'assess the applicability of a locally developed transmittance based MSI system to quantify the level of adulteration and repeated use of coconut oil'. Furthermore, the developed system in combination with the image processing algorithms was used to determine the 'level of reuse of coconut oil'. The MSI system consists of nine off-the-shelf narrowband LEDs corresponding to 9 spectral bands ranging from 375 nm to 1000 nm. Due to the less complex and cost-effective nature of the system, it is ideally suited for field deployment. Also, due to the less complexity, the equipment can be operated with minimum training. Unlike the existing equipment such as FTIR which requires more post-processing, the equipment can be configured to output the end result, such as adulteration level, which does not require further processing. Therefore, the equipment can be used to obtain real-time results. However, the acquired images carry less volume of spectral data which will increase the complexity of the system in the algorithmic section. Hence, a multi-stage algorithm is proposed in developing a model to estimate the adulteration level of coconut oil. The multi-stage algorithm includes an image preprocessing step, a Fisher Discriminant Analysis (FDA) based dimension reduction step, followed by the development of a functional relationship between Bhattacharyya distance and adulteration level. A similar multi-stage procedure was followed in determining the effect of 'reuse' on the quality of coconut oil. The algorithm includes an image preprocessing step, an FDA based dimension reduction step, and finally the development of a classifier based on spectral clustering. Therefore, the major contributions of this paper are as follows; 1. In-house development of a transmittance based MSI system 2. Introduction of a multi-stage signal processing algorithm to estimate the adulteration level of coconut oil using low-quality multispectral images. 3. Introduction of a classifier based on spectral clustering to determine the effect of reheating on the quality of coconut oil.

## II. MATERIALS AND METHODS

### A. *Preparation of Samples*

Authentic, freshly expelled coconut oil was obtained from a reputed coconut oil producers and exporters in the country, while palm oil bearing a reputed brand name was purchased from a supermarket. Samples were prepared by mixing palm oil with coconut oil at the different levels: 0, 5, 10, 15, 20, 25, 30, 35, 40 % ( $V_{palm\ oil}/V_{total}$ %) and replicated for 15 realizations. Fig. 1–(a) shows the RGB images for one set of the replicates prepared. Another 16 samples were prepared with known adulteration levels of 2, 4, 6 ,8 ,12 ,14, 16, 18, 22, 24, 26, 28, 32, 34, 36, 38 %( $V_{palm\ oil}/V_{total}$ % ) for the validation of the proposed method.

For the study on repeated use of coconut oil (see section 2.5 for a detailed analysis) potatoes were obtained from the local market, washed thoroughly with water to remove soil, peeled off, and sliced into pieces with a diameter of 4.5 cm and a thickness of 0.5 cm. The potato slices were blanched at 80°C for 1 min, blotted with a paper towel and packed in polythene bags and stored at $^-$18 °C until analysis. For each day of the experiment, a batch of potato chips was, defrosted and the drip was blotted out before the experiment. On the first day, 1L of coconut oil was heated to $170^0$C and the temperature was maintained for 10 min. Then, potato slices (100 g) were fried for 3 min at a constant temperature range of $140^0$-$150^0$C. The fried potato slices were removed from the fryer and left to drain. An image of repeatedly heated and reused coconut oil was acquired after allowing oil samples to attain the ambient temperature using the multispectral image acquisition system. The image acquisition was repeated three more times. The oil was stored for use on the following day, and the complete process was repeated for five consecutive days. Fig. 1-(b) shows the RGB images of the coconut oil samples obtained for the five days.

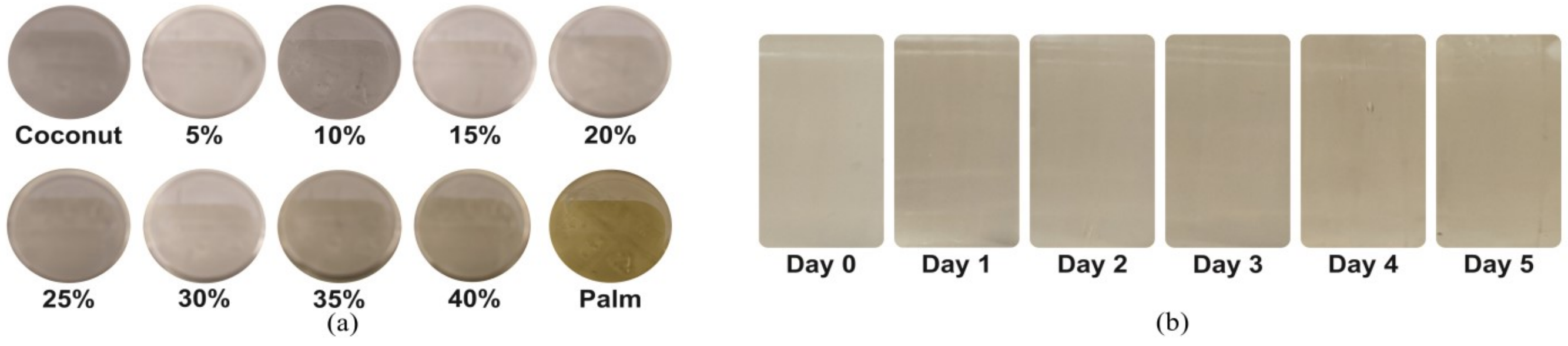

Fig. 1. RGB image of the samples prepared: (a) one set of replicates of the prepared sample for palm oil adulteration, (b) samples obtained for repeated use of reheated coconut oil

**Table 1.**: Details of the LEDs used in the LED switching circuit

| LED Number | Part number (manufacturer) | Dominant wavelength (nm) | Emitting spectral band(nm) | Half power bandwidth(nm) |
|---|---|---|---|---|
| 1 | VLMU3100 (*Vishay*) | 405 | 375 – 425 | 10 |
| 2 | SM0603BWC (*Bivar*) | 430 | 385 - 525 | 50 |
| 3 | SM1204PGC (*Bivar*) | 505 | 450 – 550 | 20 |
| 4 | 5973209202F-ND (*Dialight*) | 590 | 520 – 620 | 10 |
| 5 | 5975112402F (*Dialight*) | 660 | 630 – 685 | 20 |
| 6 | QBHP684-IR4BU (*QT Brightek*) | 740 | 690 – 760 | 20 |
| 7 | VSMY2850G (*Vishay*) | 850 | 825 – 875 | 10 |
| 8 | VSMF4710-GS08 (*Vishay*) | 890 | 865 – 915 | 10 |
| 9 | VSMS3700-GS08 (*Vishay*) | 950 | 915 - 1000 | 20 |

## *B. Multispectral Imaging Spectrum*

Most of the multispectral imagery work reported in the literature are based on the reflective properties of opaque materials. However, this method is at best only loosely applicable for most liquids, as only a small fraction of light is reflected. A low-cost multispectral imaging system (Goel et al., 2015; Prabhath et al., 2019) to measure the transmittance spectrum of liquids was developed. This imaging system has the capability of capturing monochrome multispectral images from ultraviolet (UV) to near-infrared (NIR), having an overall resolution of 9 spectral bands. The details of the LEDs which were used for this build are given in Table 1.

The imaging system used in this study consists of several major components (Fig. 2). A 10-bit CMOS monochrome camera (FLIR Blackfly S Mono,1.3 MP, USB3 Vision camera, Resolution – 1280×1024) was mounted on top of the portable dark chamber to capture the transmittance spectrum of a sample. This camera is capable of capturing images in the spectral range from 350 nm to 1080 nm. A Laptop (MSI GE62 6QD) was used to acquire the image and to send commands to both the discovery board (STM32F0DISCOVERY) and the monochrome camera. The portable dark chamber was equipped with a LED switching circuit consisting of nine off-the-shelf LEDs. An integrating hemisphere made of Aluminum having 130mm inner diameter was used to provide better illumination for the sample. A locally developed AC regulated 12V DC power supply unit was used to provide stable power input to the LED driver ICs (MAX16839ASA+). The camera was mounted in line with the LED switching circuit. The camera was tuned properly to obtain a well-focused image and the camera aperture was set to obtain an image without any saturated pixels. Two separate USB ports were used to connect the camera and the discovery board to the laptop computer. A cylindrical container was used to contain the liquid sample. A PVC pipe was used as the wall and 2 mm plain glass was used as the base of the container.

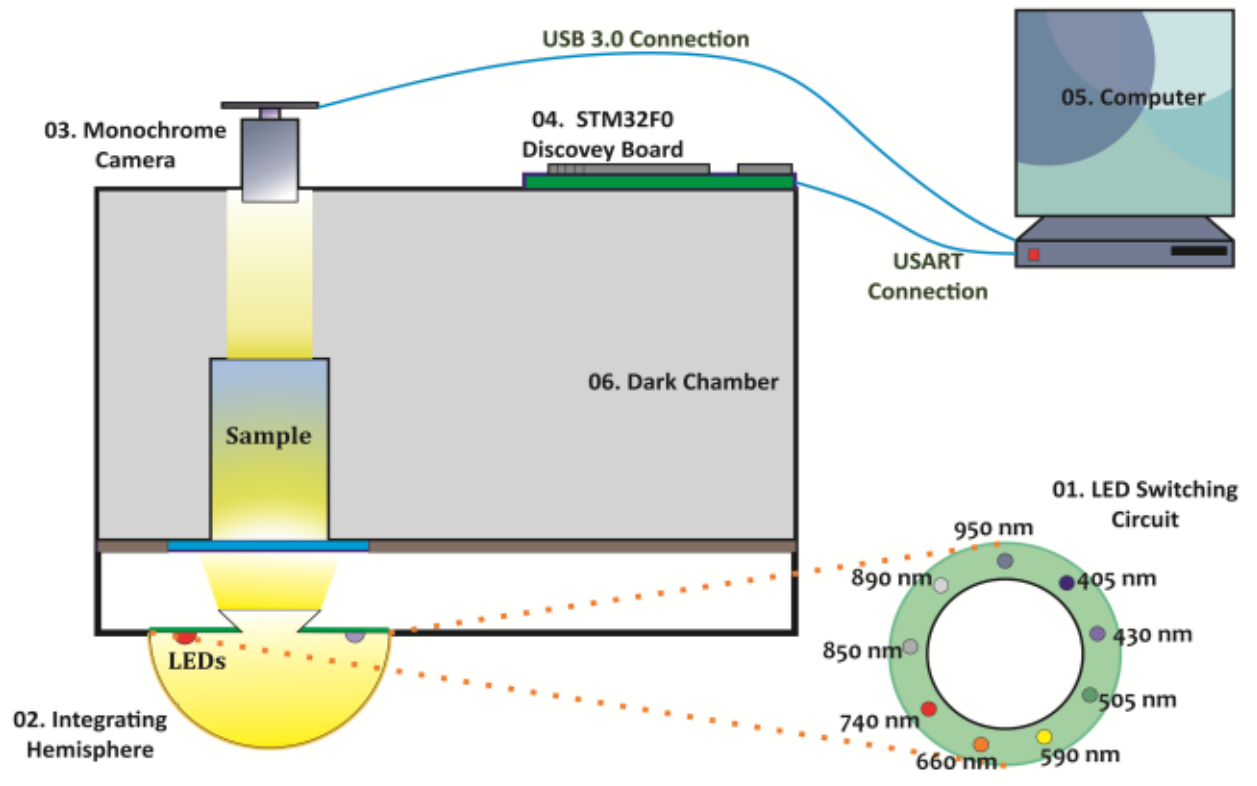

Fig. 2. Schematic diagram of the in house developed transmittance based multispectral imaging system

The image capturing process can be described as follows. Initially, the camera and the discovery board were connected to the laptop computer with separate USB cables. Then, the DC power supply was switched on. In this setup, synchronization between the camera and the LED switching circuit is important because it is required to capture the image once a LED is switched on. To maintain the synchronization, a windows batch script was used. The batch script sends a command through a USART interface to the discovery board to turn the first LED on. Once the control signal is received by the discovery board, the board sends a control signal to the relevant LED driver IC and it turns the first LED in the LED switching circuit on. Afterward, the batch script sends a command to the camera to capture the image and store it on the computer. Finally, the board sends a signal to turn off the LED. This process is repeated to all the LEDs.

## *C. Image Preprocessing*

CMOS cameras are prone to different types of noise phenomena. Some of the well-known sources of noise are readout noise and shot noise (Reibel et al., 2003). Other than that, various noises are also accumulated on the detectors due to system imperfections, such as quantization (analog to digital conversion) noise, amplifier noise, non-uniform illumination, sensitivity variations of the detectors, and dust (Qin et al., 2013).

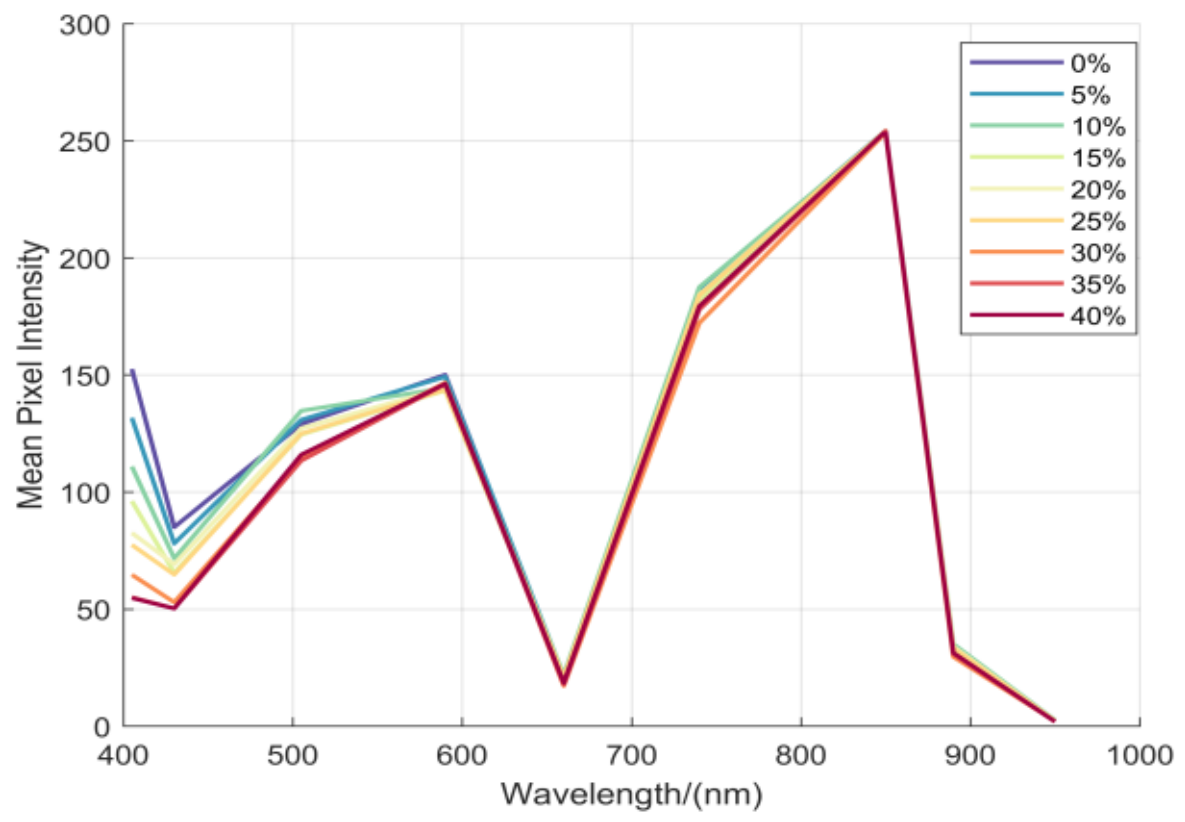

Fig. 3. Mean spectral signatures for different adulteration levels of palm oil ranging from 0% to 40%

Due to the effect of the various kinds of noises on the images, it is required to carry out several image processing steps to mitigate the noise effects. The image should undergo the process of the dark current reduction step as the first processing step before any improvements are carried out. The standard procedure for dark current reduction is: first capture an image with a closed shutter or in a dark environment (usually known as dark frames / dark current images) and then the dark frame is subtracted from the corresponding actual row images (Porter et al., 2008). In our implementation, the dark frames were captured at the beginning of each multispectral image acquisition process. Then the dark current subtraction process was applied to subsequent multispectral images utilizing the following equation,

$$P[\lambda] = S[\lambda] - D \quad (1)$$

where, $P[\lambda]$ is the dark current removed image at wavelength $\lambda$, $S[\lambda]$ is the raw image at wavelength $\lambda$ and D is the dark current image (or dark frame).

To remove the aforementioned random noise from the images, the nonlinear median filtering process was carried out on the dark current subtracted images. Here the median filter was utilized to remove some inherent noise by removing the isolated pixels while preserving the spatial resolution (Acharya and Ray, 2005). The input pixel value is replaced by the moving average filter output given by the equation,

$$P^*[i,j] = \frac{1}{N}\sum_{k=-w}^{w}\sum_{l=-w}^{w} P[i+k, j+l] \quad (2)$$

where, $P^*[i,j]$ is the replaced value of the pixel $i,j$ , $P[i,j]$ is the pixel value of the dark current subtracted image at $i,j$ , $w$ is the suitably chosen window size and $N$ is the number of pixels in the window. A window size of $30 \times 30$ is chosen for this application.

## D. *Determination of level of adulteration of coconut oil*

### i. *Spectral Signatures*

The spectral signatures of the samples were computed using the images obtained after the preprocessing step. To compute the spectral signatures, the data matrix of each adulteration level was computed using the following procedure. First, 900 pixels were selected in a 30×30 window of the multispectral image by manual inspection. The same window was selected for all the adulteration levels. As 15 replicates were prepared for each adulteration level, a total of 13500 rows (where row count, m = 13,500) were included in the data matrix, where each row represents the signatures of each of the 900 pixels, selected over each of the 15 replicate sets. The n columns of the data matrix represent the 9 spectral bands of the multispectral imaging system. Each element of the data matrix represents the spectral intensity of a pixel of the corresponding band, which is a value ranging from 0-255. Then, the data matrix was represented as X_P, for the sample with 5p% adulterated with palm oil where p = 0, 1….,8. Therefore, the dimensionality of the matrix X_P is m × n=13500 ×9. Next, all the data matrices corresponding to the nine samples were computed. After that, the mean of the 13500 pixels was calculated along the 9 spectral bands to obtain the mean spectral signature. Next, the mean spectral signature of each adulteration level was computed and plotted against the wavelength as shown in Fig. 3

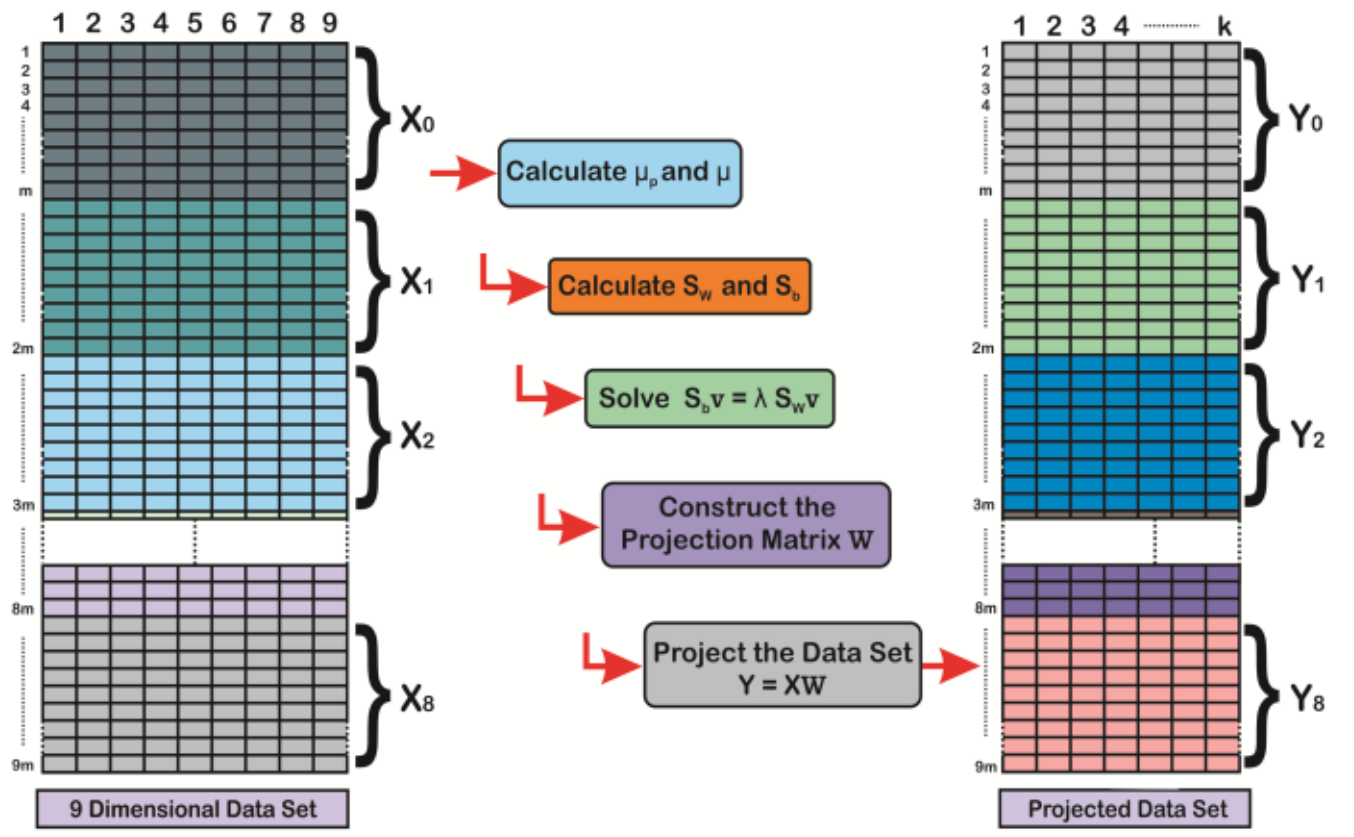

4. Overview of the FDA step: The step by step method of FDA application is depicted.

### ii. *Fisher Discriminant Analysis (FDA)*

As can be seen from Fig. 3, although there is a noticeable difference in the 400 nm band, the difference is less significant in the other bands. Therefore, to increase the separability of classes and to reduce data redundancy, Fisher Discriminant Analysis (FDA) (Mika et al., 1999) was performed on the nine-dimensional data set.

First, data matrix $X$ was constructed by combining the data matrix of each adulteration level as, $X = [X_0\, X_1\, X_2 X_3\, X_4\, X_5\, X_6\, X_7\, X_8\,]^T$ where $T$ represents the matrix transpose. Therefore, the dimensionality of $X$ is $9m \times n = 121500 \times 9$, where the rows denote the total number of pixels selected from each multispectral image and columns denote the nine spectral bands. After that, FDA was performed on the data matrix $X$ in the following procedure (Fig. 4 represents the overview of the procedure).

- Step 1: Compute the mean vector $\mu_p$ of each class and the total mean vector $\mu$ (dimensions = 1 × 9 ):

$$\mu_p = \frac{1}{m}\sum_{i=1}^{m} X_p(i) \quad (3)$$

$$\mu = \frac{1}{9m}\sum_{i=1}^{9m} X(i) \quad (4)$$

- Step 2: Compute within the class scatter matrix $S_w$ (dimensions = 9 × 9).

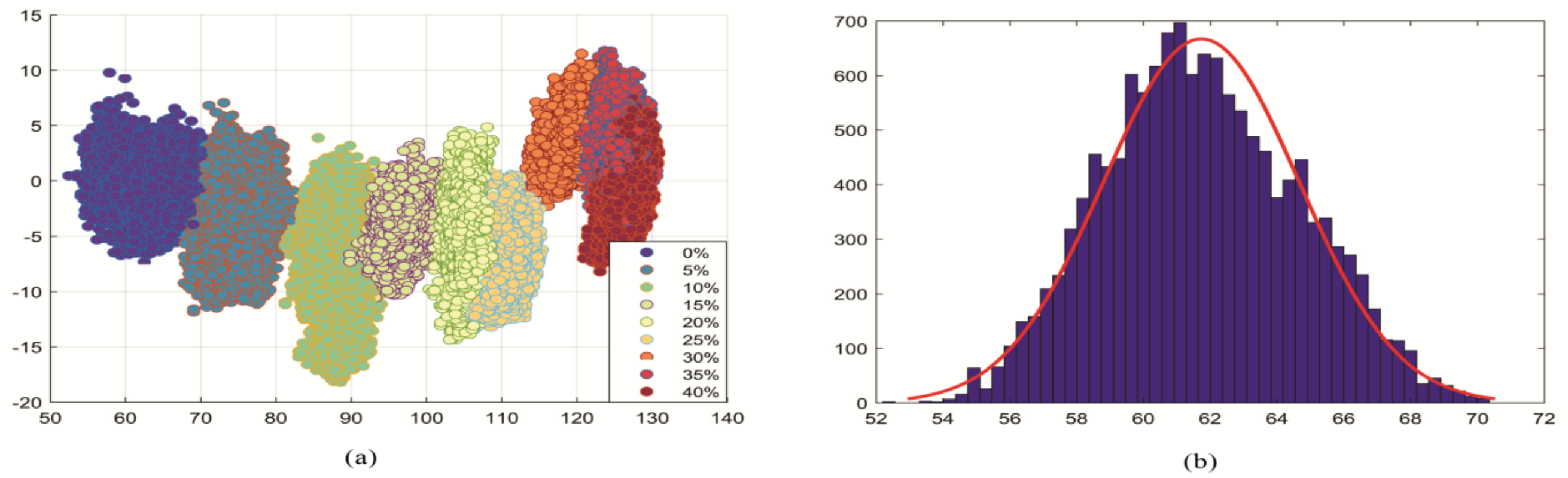

Fig. 5. Resultant dataset after FDA application: (a) Reduced dataset plotted along the first two eigenvectors, (b) Approximate Gaussian distribution of the reduced dataset along the first eigen vector

$$S_w = \sum_{i=1}^{9} C_i \quad (5)$$

$$C_p = \sum_{i=1}^{m} (X_p(i) - \mu_p)(X_p(i) - \mu_p)^T \quad (6)$$

where, $C_p$ is the covariance matrix of the p% adulterated class.

- Step 3: Compute between the class scatter matrix $S_b$ (dimensions = 9 × 9).

$$S_b = \sum_{i=1}^{9} (\mu_i - \mu)(\mu_i - \mu)^T \quad (7)$$

- Step 4: Solve the generalized eigenvalue problem,

$$S_b v = \lambda S_w v \quad (8)$$

- Step 5: Extract the eigenvectors corresponding to the highest eigenvalues $k\ (\leq 9)$ and construct the projection matrix $W$ (dimensions = $9 \times k$).

$$W = [v_1, v_2, \ldots, v_k] \quad (9)$$

- Step 6: Project the data set into the reduced $k$ dimensional data space $Y$ (dimensions = $121500 \times k$).

$$Y = XW \quad (10)$$

The value of $k$ was selected to be five to reduce the dimensionality of the data space while minimizing the information loss by retaining more than 99% of the total variance. The nine-dimensional raw data set was reduced to a five-dimensional data set. The resulting data set is shown in Fig. 5-(a) plotted along the first two eigenvectors.

*iii. Model Construction*

To establish a functional relationship with the adulteration level, Bhattacharya distance (Kailath, 1967) was utilized. This metric measures the similarity between 2 Gaussian distributions. The spectral intensity values of each adulteration level were observed along the eigenvectors and they formed an approximate Gaussian distribution as shown in Fig. 5-(b). Hence, the probability distribution corresponding to each adulteration level was assumed to be a five-dimensional multivariate gaussian distribution. The 5 × 1 dimensional mean vector and 5 × 5 dimensional covariance matrix were obtained from the resultant dataset to which FDA was applied to. For a multivariate Gaussian distribution, Bhattacharyya distance can be computed as,

$$B(f_{y0}, f_{yp}) = \frac{1}{8}(\mu_{y0} - \mu_{yp})^T \left(\frac{C_{y0} - C_{yp}}{2}\right)^{-1} (\mu_{y0} - \mu_{yp}) + \frac{1}{2} ln \left[\frac{det\, \frac{C_{y0} - C_{yp}}{2}}{\sqrt{det\,(C_{y0}) det\,(C_{yp})}}\right], \quad (11)$$

where, $B(f_{y0}, f_{yp})$ is the Bhattacharyya distance between the two multivariate Gaussian distribution functions $f_{y0}$ and $f_{yp}$ , $\mu_{y0}\mu_{yp}$, are the mean vectors and $C_{y0}$, $C_{yp}$ are the covariance matrices of classes $Y_0$ and $Y_P$, respectively. Class $Y_0$ represents the class of pure coconut oil and was used as the reference class in computing the above metrics. Class $Y_P$represents $5p$% adulterated coconut oil where $p = 0,1,\ldots,8$. $T$, $(\cdot)^{-1}$ and $det$ represent matrix transpose, matrix inverse, and matrix determinant, respectively. The accuracy of the predicted model and its validation and verification is discussed in the Results and Discussion section.

## *E. Study on repeated use of coconut oil*

The reheating of coconut oil introduces thermophysical changes (Bhuiyan et al., 2016) to the oil which are reflected in the spectral signature of the oil. Therefore, the difference between repeatedly heated oil and pure or once heated oil was visible to the naked eye. However, determining the number of times or the duration the oil was heated, is impossible to the naked eye, let alone a qualitative analysis. In this section, an algorithm is proposed to cluster the heated oil into qualitative groups from an ensemble. The proposed algorithm is then validated through emulated field tests.

*i. Model Construction*

Multispectral images of the oil samples were obtained after each heating cycle up to 5 cycles. Each monochrome image was $30 \times 30$ square pixels in size. With a number of different trials, an abundance of data was recorded for each heating cycle. From this dataset, a number of data points were selected randomly at uniform intervals in each image. This reduces the computational complexity and the dependence on the training dataset.

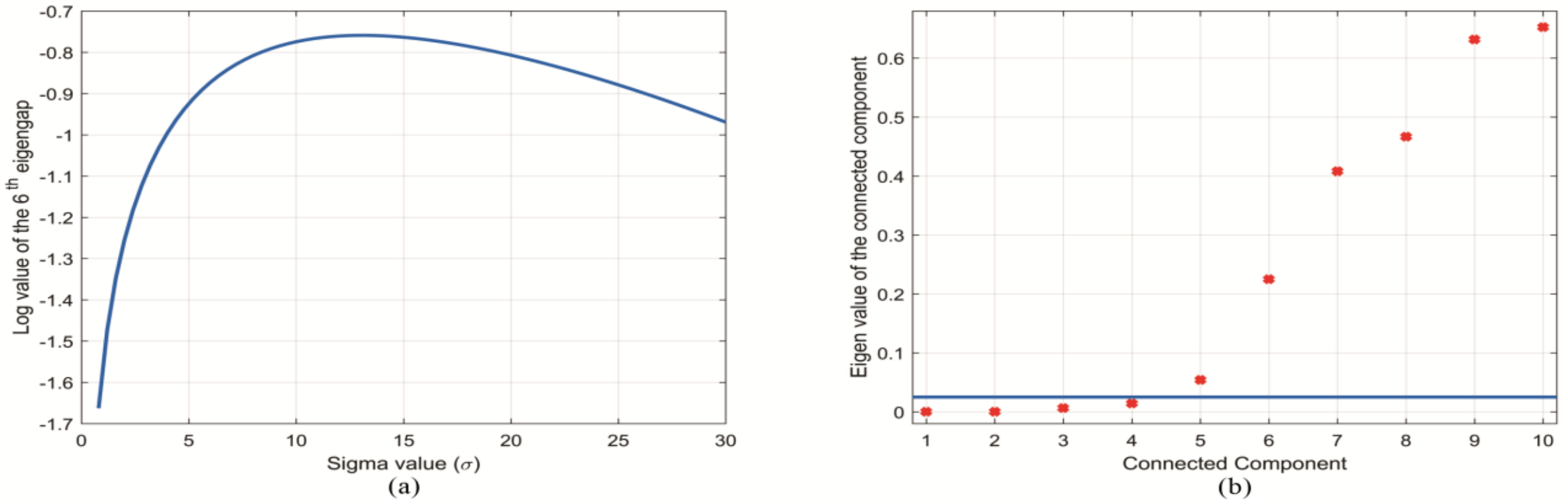

Fig. 6. Application of sigma sweep: (a) Variation of sixth eigengap with σ, (b) Variation of the eigenvalue of connected components

*ii. Constructing the classifier*

The purpose of the solution is to group, tested oils, into different qualitative classes. Hence, a classifier was sufficient rather than a functional relationship since the variation is discrete. In order to construct the classifier, first, the dimensionality of the dataset was reduced. This was performed using FDA for the training dataset including samples from pure oil as well. This algorithm produces a transformation matrix from the original basis to a lower-dimensional basis. In this reduced space, the separation amongst different classes is increased and this transformation constructs a frame of reference for the classifier. The reference cluster of this lower-dimensional space is the pure oil class, as its spectral properties are unaffected.

Next, spectral clustering was performed to identify the number of qualitative classes present in the ensemble. Spectral clustering (Ng et al., 2001) is an unsupervised classification algorithm. This algorithm assesses the spectral connectivity of the multispectral dataset. Hence, the spectral connectivity of the dataset has to be improved for optimum classification. This optimization was performed using the sigma sweep method (Rupasinghe et al., 2016) for the six heat classes, this in return will further increase the connectivity for the quantitative

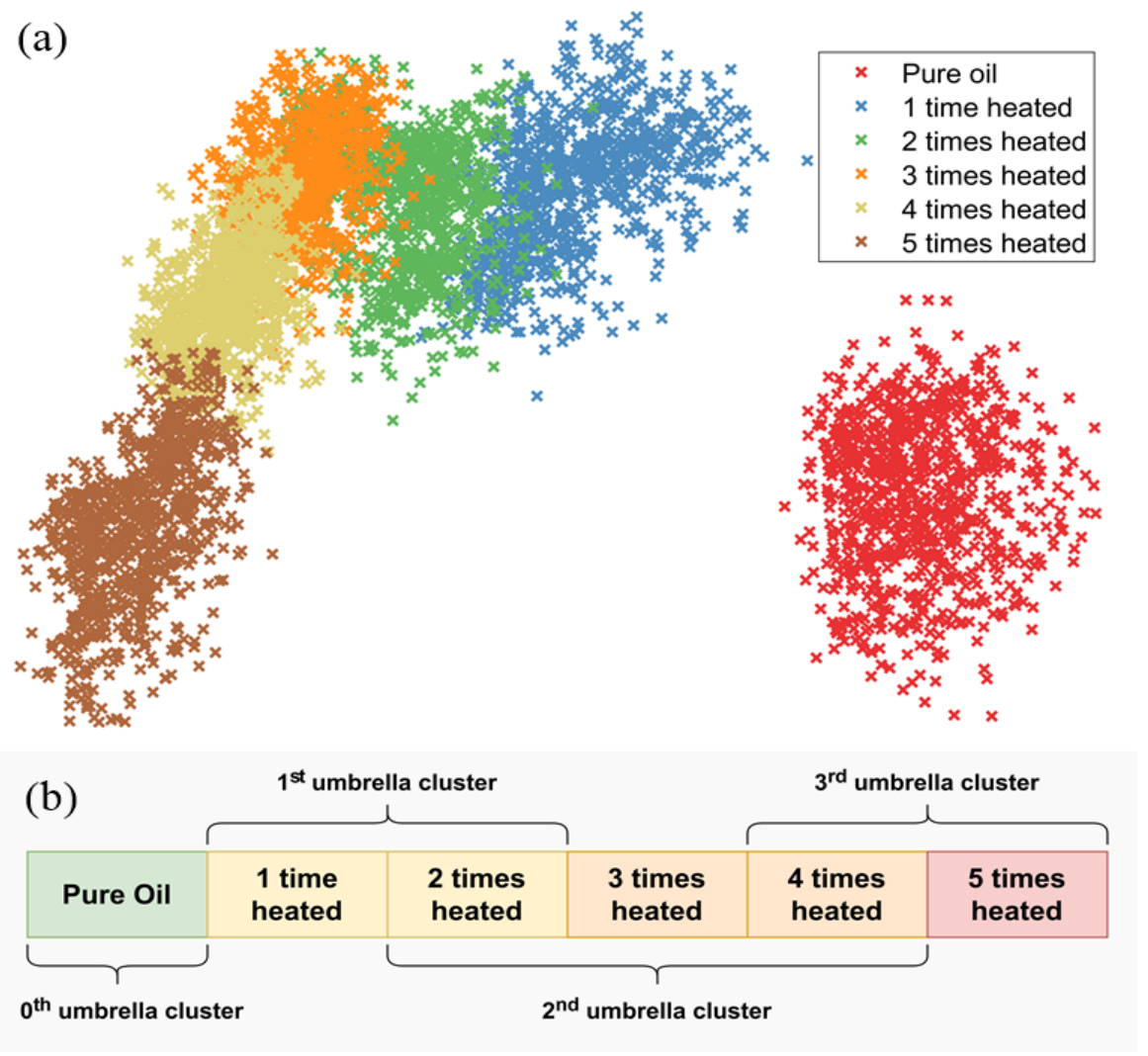

Fig. 7. Classification of training dataset: (a) Low dimensional representation of the training dataset, (b) Qualitative classes with respective heat classes

analysis. The matrices required for the algorithm were computed according to the equations,

$$W(i,j) = \begin{cases} exp^{-\frac{\|x_i - x_j\|^2}{2\sigma^2}} & ; i \neq j, \\ 0 & \end{cases} \quad (12)$$

$$D(i,i) = \sum_j W(i,j), \quad (13)$$

$$L = I - D^{-\frac{1}{2}} W D^{-\frac{1}{2}}, \quad (14)$$

where, $W$ is the affinity matrix, $D$ is the degree matrix, $L$ is the Graph Laplacian and $I$ is the identity matrix.

First, $W$ of the graph was constructed using equation 12, using a Gaussian kernel, where $x_i, x_j$ and $\sigma$ are, the $i^{th}$pixel vector, $j^{th}$ pixel vector and sigma value, respectively. Then $D$ was constructed according to equation 13 whose diagonal entries are the cumulative affinities of the corresponding pixel. Finally, according to equation 14, the normalized Laplacian of the graph was computed. Then an eigen decomposition was performed on $L$ to find the eigenvalue and the eigenvector of each connected component. Since the task was set to optimize $\sigma$ for six classes, the difference between the $5^{th}$ and the $6^{th}$ eigenvalue known as the $6^{th}$ eigengap was considered. Then the variation of the eigengap with the $\sigma$ was plotted on a semi-logarithmic graph as shown in Fig.6-(a) to analyze.

Thereafter, the $\sigma$, for which the highest eigengap value was recorded, was set as the optimum value for the training dataset. Accordingly, equation 12 to equation 14 were recomputed for the optimum $\sigma$ value and an eigen decomposition was performed. Afterward, the first six eigenvectors corresponding to the six least significant eigenvalues of the positive-semi definite matrix $L$ were considered for the construction of the projection matrix which projects the data into a new Euclidean space, where the spectral connectivity is now improved. In this matrix, each row corresponds to a pixel from the original dataset and the columns contain the coordinates of each pixel in this new space. Next, the newly formed dataset was clustered using K-means (Macqueen, 1967) algorithm with random initial conditions.

To develop the qualitative class classifier, the number of disconnected components in the graph was required. For this, the eigenvalues of the $L$ were considered. Theoretically, the eigenvalue of each connected component is 0, but when the Gaussian kernel is used the graph is treated as a fully connected graph. Hence, the eigenvalues closer to 0 would give the number of connected components. A threshold of 0.025 was used to count the number of connected components. Since the normalized graph Laplacian is used, the eigenvalues are in the range of 0 to 1; the threshold is not dependent on the dataset. As shown in Fig. 6-(b) there are four connected components in the graph with eigenvalues below the threshold, hence, the number of qualitative classes was set to be four.

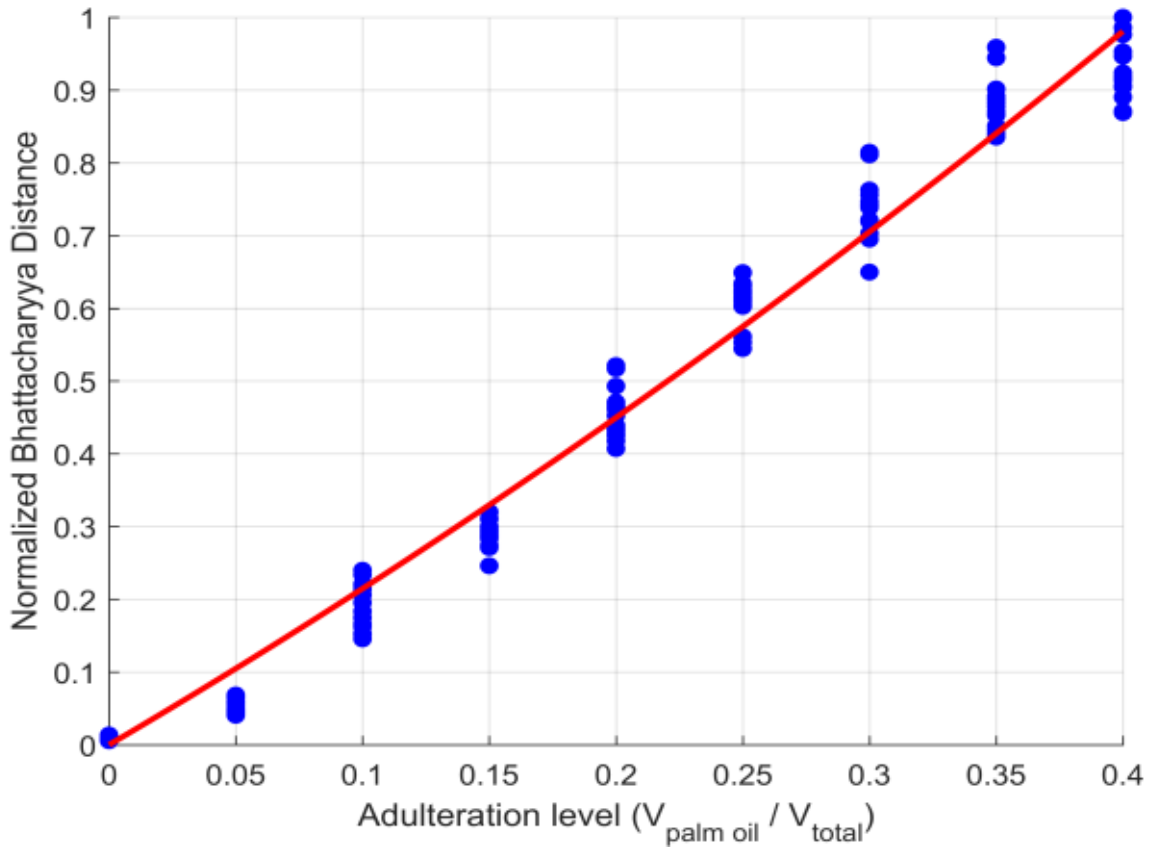

Fig. 8 Development of a functional relationship using statistical a metric: The relationship between the quantity of palm oil mixed with coconut oil and normalized Bhattacharyya distance

In the qualitative analysis, the clusters identified through spectral clustering followed up by K-means, should be assigned to qualitative classes. For this, the distance to the center of the cluster from the center of the reference cluster was used. If the distance is within a certain range, then the cluster will be assigned to the qualitative class for which that range was defined. Before, calculating those predefined distance measurements, it was necessary to decide which heat classes are included in those four qualitative classes.

Since the proposed algorithm performs a qualitative analysis of the dataset, the relative behavior of those heat classes was considered rather than assigning a heat class to a particular qualitative class. As shown in Fig. 7-(a) there are overlaps between certain heat classes which validates the above argument. Let us consider the heat classes excluding the pure oil class. The first two classes, the last two classes, and the three intermediate classes will construct a qualitative class each and the pure oil class will construct a separate qualitative class as shown in Fig. 7-(b). Then, the distance to the class mean for each umbrella cluster was calculated using the training dataset.

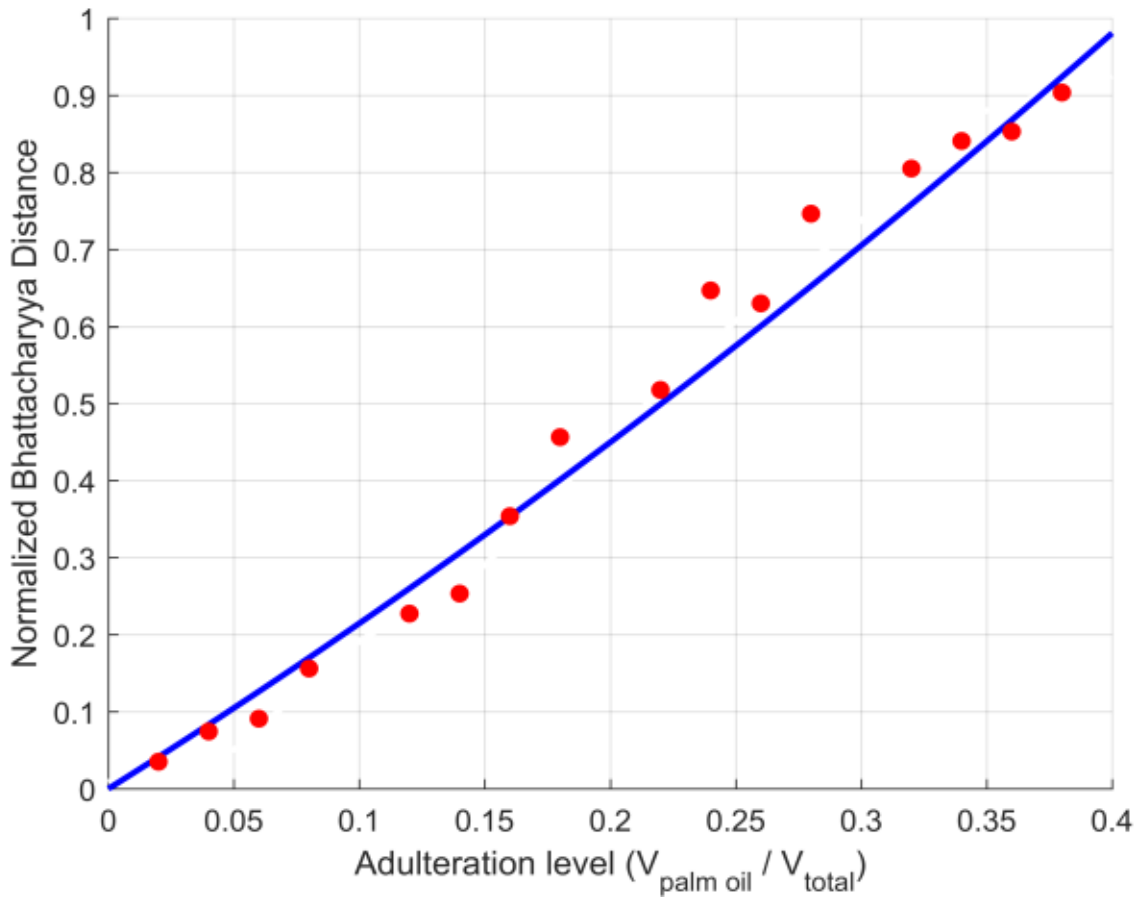

Fig. 9 Validation of the proposed algorithm for palm oil adulteration. Validation samples closely follow the functional relationship with an MSE of 0.0029

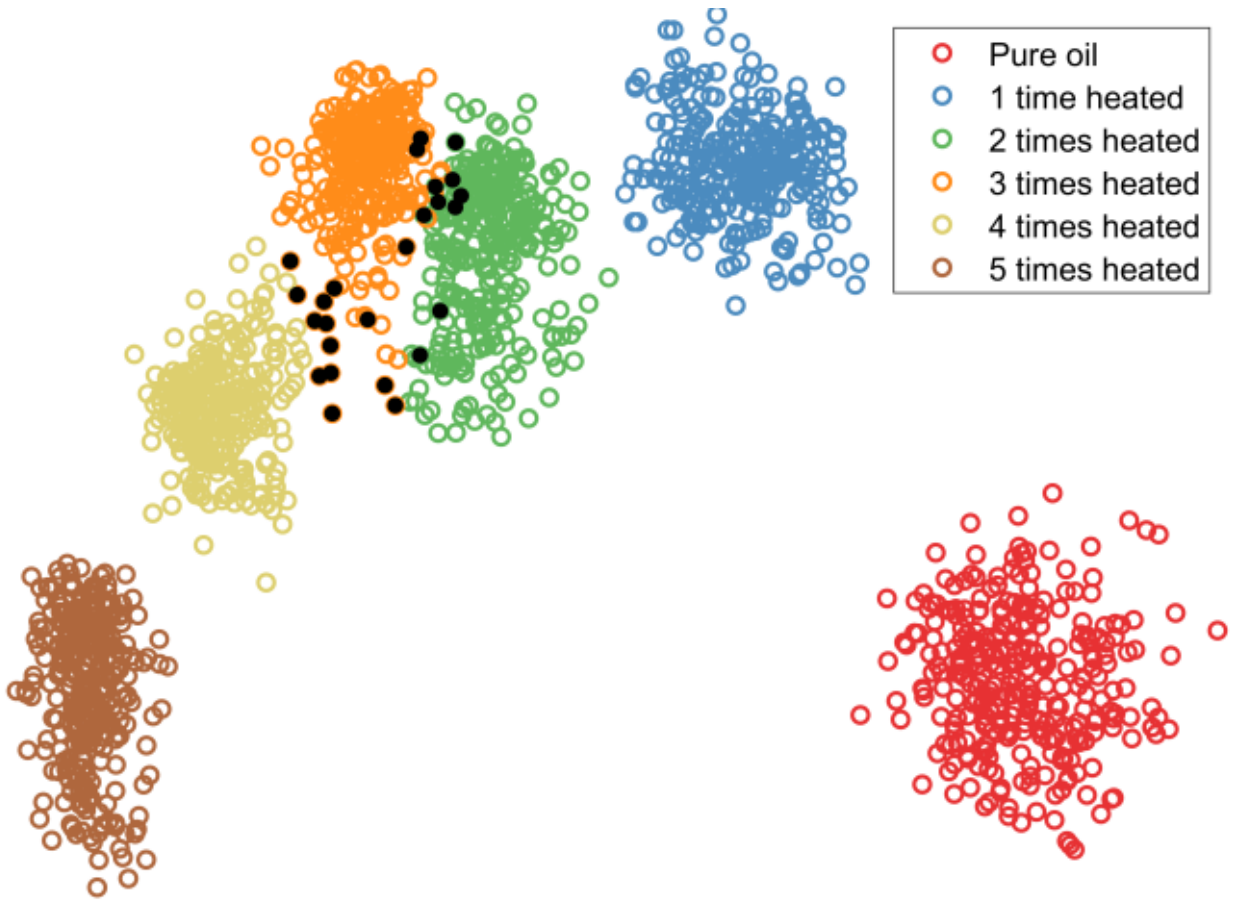

Fig. 10 Six-class classification of repeatedly used coconut oil

## III. Results and Discussion

### A. Determination of the level of adulteration of coconut oil

A functional relationship was established between Bhattacharyya distance and adulteration level as shown in Fig. 8. The above metric was calculated using equation 11 for all 15 replicates at each adulteration level and plotted as shown in Fig. 8. Nine hundred data points were randomly selected from the class $Y_0$ and used as the reference when calculating the above metrics. The mean of the 15 replicates was selected in generating a functional relationship with the adulteration level. The generated functional relationship for Normalized Bhattacharyya distance was given by, $Y = 1.016X^2 + 2.045X$ with $R^2 = 0.9876$. Here, $X$ denotes the percentage adulteration level.

#### *i. Validation of the model*

The proposed algorithm which utilizes Bhattacharyya distance for the estimation of adulteration level has been validated through 16 independent samples prepared with known adulteration levels as mentioned in section 2.1. Therefore, it can be noted that the established ground truth was known accurately as the validation samples were prepared with precise volume measurements. The adulteration level of each sample was estimated using the proposed algorithm and plotted in the calibrated model as shown in Fig. 9. It can be clearly seen that the error between the actual adulteration level and the estimated adulteration level is very low with an MSE of 0.0029. Hence, it can be assumed that the goal of estimating the adulteration level from the proposed method was a success.

### B. Study on repeated use of coconut oil

#### *i. Validation of the algorithm*

The performance of the classifier was evaluated for both training and validation datasets, accordingly. First, the datasets were created through random selection of pixels from each repeatedly used oil class. Due to this random selection process, five different validation datasets were created along with their respective ground truths. Then, a classifier was developed, several times, for different training sample sizes, and; the accuracy of the classifier for each training sample size was recorded. Then, the classifier with the highest accuracy, recorded during the training phase was used for the validation process. For the five validation datasets, the accuracy and the repeatability of the results were considered to evaluate the performance of the classifier.

**Table 2**: Accuracy and repeatability of results for the trials

| Description | Heat Class Classification | | | Qualitative Class Classification | | | |
|---|---|---|---|---|---|---|---|
| | Max accu. | Min accu. | Mode accu. | Max accu. | Min accu. | Mode accu. | ReSc |
| TS1 | 100.00 | 78.33 | 92.50 | 100.00 | 79.44 | 97.00 | 0.97 |
| TS2 | 86.89 | 54.33 | 70.00 | 100.00 | 97.75 | 99.50 | 0.98 |
| TS3 | 75.67 | 53.33 | 65.90 | 99.70 | 96.70 | 98.50 | 0.98 |

The accuracy of the algorithm varied between 85% and 99% for the five validation datasets. In Fig. 10 the most repeated result is shown and the accuracy of that is 97%. Next, the qualitative analysis of the validation datasets was performed.

In the result shown in Fig. 11-(a) for one of the datasets the classifier has separated data into four qualitative umbrella clusters. The qualitative analysis produced results that were more than 90% accurate. The result shown here has a mode accuracy of 98.5% recorded for 20 trials.

*ii. Emulated field test*

To perform the field test three oil samples were prepared. The testing datasets were constructed with randomly selected pixels in the multispectral image of each sample. Therefore, five datasets were created for each sample to test the repeatability of the classification and the accuracy of results for each dataset was recorded for 30 trials. A repeatability score was defined to quantify the repeatability of the classifier as,

$$ReSc = 1 - {LS}/{MS} \qquad (15)$$

where, $ReSc$ is the repeatability score, $LS$ is the least classified times to a qualitative class and $MS$ is the most classified times to a qualitative class. The repeatability score of a dataset was taken as the average of the individual scores for each trial of a dataset and the average of all datasets was the score of the corresponding sample.

The results for the qualitative analysis are displayed in Fig. 11 (b), (c), (d) for the three samples. The number of heat classes included in each qualitative class is different among the samples. However, the classification of classes was in

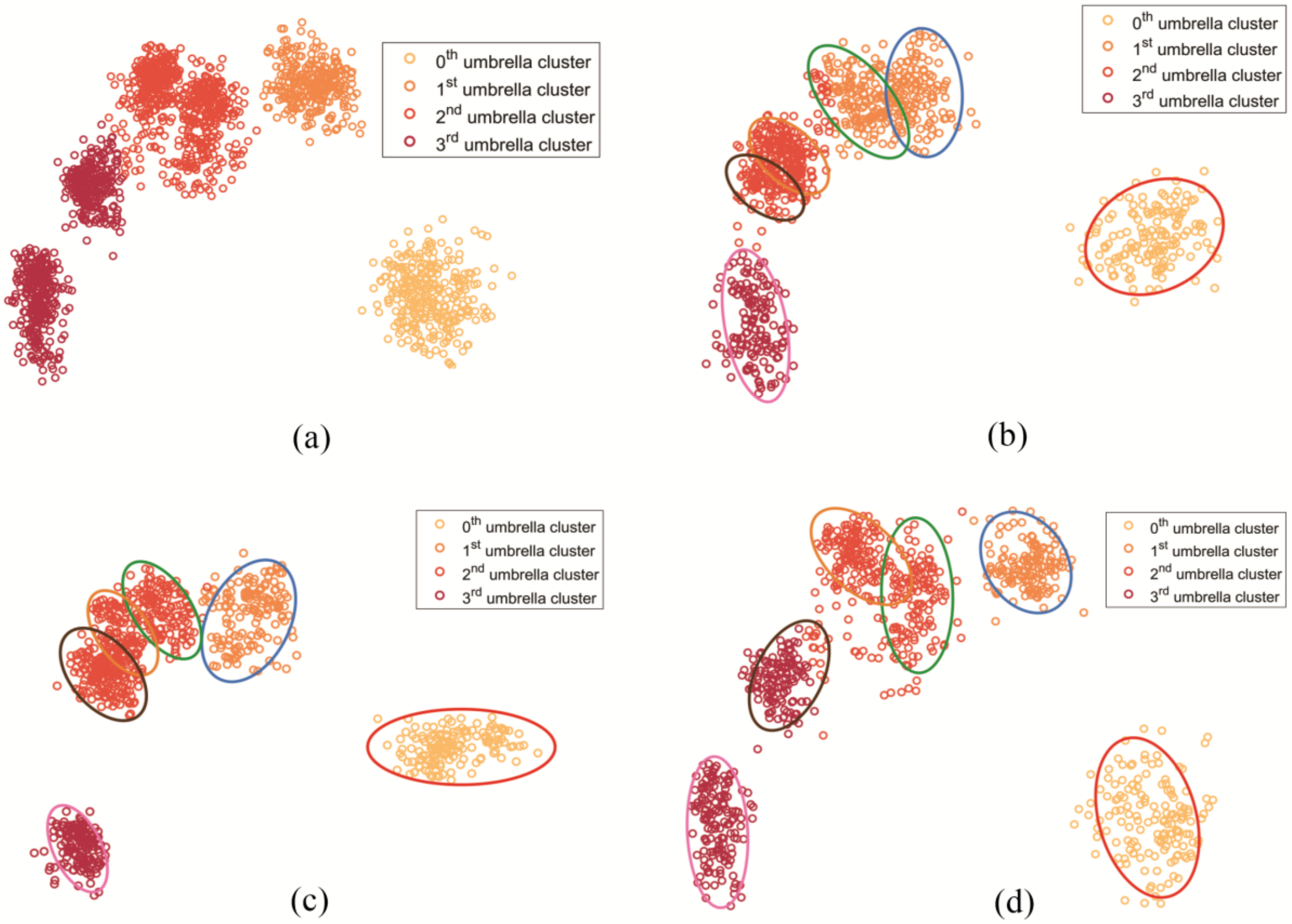

Fig. 11Classification results for the repeatedly used coconut oil: (a) Qualitative classification of repeatedly used coconut oil (b) Qualitative classification with 2 times heated class under 2nd umbrella cluster and 4 times heated class under 3rd umbrella cluster (c) Qualitative classification with 2 times heated and 4 times heated classes under 2nd umbrella cluster (d) Qualitative classification with 2 times heated class under 1st umbrella cluster and 4 times heated class under 2nd umbrella cluster accordance with the class arrangement proposed in Fig. 7-(b). This is a clear indication regarding the capability of the algorithm to assess the changes in the oil properties, rather than simply separating into groups based on the number of reheats.
In Table 2, the results for the heat class classification and the qualitative class classification are presented. The heat class classification accuracy is higher when the separation is greater among the classes. Due to this reason, the mode accuracy and the accuracy range was different for the three oil samples. Interestingly, for the qualitative analysis, the difference between these two measures among the three classes was comparatively low. This indicates the strength of the classifier in qualitative analysis. Along with the accuracy, the repeatability of the results should be evaluated. The $ReSc$ values for the three samples were between 4.85 and 4.91. This suggests that the algorithm produces the same result most of the time. When we consider the class, classification proposed in Fig. 11, the $ReSc$ is sensitive to the variations in the 2-time heated and the 4-time heated classes. In an ideal situation, spectral signatures of the same heat class should not have a huge variation. On that note, when the $ReSc$ for a sample is less than the perfect score of 5.0, it indicates that there are subtle differences in the qualitative aspects of the sample. The classifier was able to identify these changes when multiple datasets were prepared and tested.
The emulated field testing showed that the changes introduced to the oil during reheating can be identified through multispectral imagery-based algorithms. Further, this supports the validity of the proposed algorithm for industrial applications.

## IV. Conclusion

This study presents two novel multi-stage signal processing algorithms to estimate the level of adulteration of authentic coconut oil, adulterated with palm oil, and a mechanism to determine the number of times a coconut oil sample has been repeatedly heated. The algorithms are developed for multispectral images acquired from an in-house developed transmittance-based multispectral imaging system. A high curve fitting ($R^2$ = 0.9876) was achieved for the adulteration level estimating algorithm whereas, high accuracies (mode accuracy $\geq$ 0.90) were recorded for the classifying algorithm. Moreover, the proposed algorithms were applied on independent samples with known adulteration levels, and number of reheats, for validation. The low mean square error of 0.0029 for adulteration level identification coupled with 0.7616 error metric for six class classification and 0.983 error metric for qualitative classification of repeatedly used oil, demonstrates a remarkable level of accuracy the proposed solution can be easily modified or extended to handle a new class of adulterants and a higher number of heating cycles.

## V. acknowledgement

Financial support for this research was provided by the University of Peradeniya, Sri Lanka research grant (Research Grant No: URG/2017/26/E).

We would like to express our special thanks to Silver Mills Group for supplying coconut oil for the experiment. We would like to express our sincere gratitude to Ms. E.G.T.S. Wijethunga of the Department of Food Science and Technology, University of Peradeniya. We would also like to show our gratitude to the Department Head of the Department of Food Science and Technology for granting permission to use the required laboratory facilities.

## References

Acharya, T., Ray, A.K., 2005. Image Processing Principles and Applications.

Achata, E.M., 2020. Visible and NIR hyperspectral imaging and chemometrics for prediction of microbial quality of beef Longissimus dorsi muscle under simulated normal and abuse storage conditions. Food Sci. Technol. 14.

Amit, Jamwal, R., Kumari, S., Kelly, S., Cannavan, A., Singh, D.K., 2020. Rapid detection of pure coconut oil adulteration with fried coconut oil using ATR-FTIR spectroscopy coupled with multivariate regression modelling. LWT 125, 109250. https://doi.org/10.1016/j.lwt.2020.109250

Bhuiyan, M.T.H., Khan, M., Rahman, A., Chowdhury, U.K., 2016. Effect of Reheating on Thermophysical Properties of Edible Oil at High Temperature. Int. J. Adv. Res. Phys. Sci. 3, 30–34.

Brosnan, T., Sun, D.-W., 2004. Improving quality inspection of food products by computer vision—a review. J. Food Eng. 61, 3–16. https://doi.org/10.1016/S0260-8774(03)00183-3

Chaminda Bandara, W.G., Kasun Prabhath, G.W., Sahan Chinthana Bandara Dissanayake, D.W., Herath, V.R., Roshan Indika Godaliyadda, G.M., Bandara Ekanayake, M.P., Demini, D., Madhujith, T., 2020. Validation of multispectral imaging for the detection of selected adulterants in turmeric samples. J. Food Eng. 266, 109700. https://doi.org/10.1016/j.jfoodeng.2019.109700

Cheng, J.-H., Sun, D.-W., 2015. Rapid and non-invasive detection of fish microbial spoilage by visible and near infrared hyperspectral imaging and multivariate analysis. LWT - Food Sci. Technol. 62, 1060–1068. https://doi.org/10.1016/j.lwt.2015.01.021

De Alzaa F, Guillaume C, Ravetti L, 2018. Evaluation of Chemical and Physical Changes in Different Commercial Oils during Heating 2.

Ebrahimi, P., van den Berg, F., Aunsbjerg, S.D., Honoré, A., Benfeldt, C., Jensen, H.M., Engelsen, S.B., 2015. Quantitative determination of mold growth and inhibition by multispectral imaging. Food Control 55, 82–89. https://doi.org/10.1016/j.foodcont.2015.01.050

ElMasry, G., Wang, N., Vigneault, C., Qiao, J., ElSayed, A., 2008. Early detection of apple bruises on different background colors using hyperspectral imaging. LWT - Food Sci. Technol. 41, 337–345. https://doi.org/10.1016/j.lwt.2007.02.022

Femenias, A., Gatius, F., Ramos, A.J., Sanchis, V., Marín, S., 2020. Standardisation of near infrared hyperspectral imaging for quantification and classification of DON contaminated wheat samples. Food Control 111, 107074. https://doi.org/10.1016/j.foodcont.2019.107074

Goel, M., Patel, S.N., Whitmire, E., Mariakakis, A., Saponas, T.S., Joshi, N., Morris, D., Guenter, B., Gavriliu, M., Borriello, G., 2015. HyperCam: hyperspectral imaging for ubiquitous computing applications, in: Proceedings of the 2015 ACM International Joint Conference on Pervasive and Ubiquitous Computing - UbiComp '15. Presented at the the 2015 ACM International Joint Conference, ACM Press, Osaka, Japan, pp. 145–156. https://doi.org/10.1145/2750858.2804282

Gowen, A., Odonnell, C., Cullen, P., Downey, G., Frias, J., 2007. Hyperspectral imaging – an emerging process analytical tool for food quality and safety control. Trends Food Sci. Technol. 18, 590–598. https://doi.org/10.1016/j.tifs.2007.06.001

Hamsi, M.A., Othman, F., Das, S., Kamisah, Y., Thent, Z.C., Qodriyah, H.M.S., Zakaria, Z., Emran, A., Subermaniam, K., Jaarin, K., 2015. Effect of consumption of fresh and heated virgin coconut oil on the blood pressure and inflammatory biomarkers: An experimental study in *Sprague Dawley* rats. Alex. J. Med. 51, 53–63. https://doi.org/10.1016/j.ajme.2014.02.002

Huang, W., Li, J., Wang, Q., Chen, L., 2015. Development of a multispectral imaging system for online detection of bruises on apples. J. Food Eng. 146, 62–71. https://doi.org/10.1016/j.jfoodeng.2014.09.002

Jamwal, R., Kumari, S., Dhaulaniya, A.S., Balan, B., Kumar, D., 2019. Application of ATR-FTIR spectroscopy along with regression modelling for the detection of adulteration of virgin coconut oil with paraffin oil 38.

Kailath, T., 1967. The Divergence and Bhattacharyya Distance Measures in Signal Selection. IEEE Trans. Commun. 15, 52–60. https://doi.org/10.1109/TCOM.1967.1089532

Kamruzzaman, M., Makino, Y., Oshita, S., 2016. Hyperspectral imaging for real-time monitoring of water holding capacity in red meat. LWT - Food Sci. Technol. 66, 685–691. https://doi.org/10.1016/j.lwt.2015.11.021

Ma, F., Yao, J., Xie, T., Liu, C., Chen, W., Chen, C., Zheng, L., 2014. Multispectral imaging for rapid and non-destructive determination of aerobic plate count (APC) in cooked pork sausages. Food Res. Int. 62, 902–908. https://doi.org/10.1016/j.foodres.2014.05.010

Ma, J., Cheng, J.-H., Sun, D.-W., Liu, D., 2019. Mapping changes in sarcoplasmatic and myofibrillar proteins in boiled pork using hyperspectral imaging with spectral processing methods. LWT 110, 338–345. https://doi.org/10.1016/j.lwt.2019.04.095

Macqueen, J., 1967. SOME METHODS FOR CLASSIFICATION AND ANALYSIS OF MULTIVARIATE OBSERVATIONS. Multivar. Obs. 1, 281–297.

Manaf, M.A., Man, Y.B.C., Hamid, N.S.A., Ismail, A., Abidin, S.Z., 2007. ANALYSIS OF ADULTERATION OF VIRGIN COCONUT OIL BY PALM KERNEL OLEIN USING FOURIER TRANSFORM INFRARED SPECTROSCOPY. J. Food Lipids 14, 111–121. https://doi.org/10.1111/j.1745-4522.2007.00066.x

Mika, S., Ratsch, G., Weston, J., Scholkopf, B., Mullers, K.R., 1999. Fisher discriminant analysis with kernels, in: Neural Networks for Signal Processing IX: Proceedings of the 1999 IEEE Signal Processing Society Workshop (Cat. No.98TH8468). Presented at the Neural Networks for Signal Processing IX: 1999 IEEE Signal Processing Society Workshop, IEEE, Madison, WI, USA, pp. 41–48. https://doi.org/10.1109/NNSP.1999.788121

Ng, A.Y., Jordan, M.I., Weiss, Y., 2001. On Spectral Clustering: Analysis and an algorithm, in: ADVANCES IN NEURAL INFORMATION PROCESSING SYSTEMS. MIT Press, pp. 849–856.

Nunes, C.A., 2014. Vibrational spectroscopy and chemometrics to assess authenticity, adulteration and intrinsic quality parameters of edible oils and fats. Food Res. Int. 60, 255–261. https://doi.org/10.1016/j.foodres.2013.08.041

Ouyang, Q., Wang, L., Park, B., Kang, R., Wang, Z., Chen, Q., Guo, Z., 2020. Assessment of matcha sensory quality using hyperspectral microscope imaging technology. LWT 125, 109254. https://doi.org/10.1016/j.lwt.2020.109254

Pandiselvam, R., Manikantan, M.R., Ramesh, S.V., Beegum, S., Mathew, A.C., 2019. Adulteration in Coconut and Virgin Coconut Oil-Implications and Detection Methods 19–22.

Porter, W.C., Kopp, B., Dunlap, J.C., Widenhorn, R., Bodegom, E., 2008. Dark current measurements in a CMOS imager, in: Blouke, M.M., Bodegom, E. (Eds.), . Presented at the Electronic Imaging 2008, San Jose, CA, p. 68160C. https://doi.org/10.1117/12.769079

Prabhath, G., Bandara, W., Dissanayake, D., Hearath, H., Godaliyadda, G., Ekanayake, M., Demini, S., Madhujith, T., 2019. Multispectral Imaging for Detection of Adulterants in Turmeric Powder, in: Hyperspectral Imaging and Sounding of the Environment. Optical Society of America, pp. HTu3B–3.

Qin, J., Chao, K., Kim, M.S., Lu, R., Burks, T.F., 2013. Hyperspectral and multispectral imaging for evaluating food safety and quality. J. Food Eng. 118, 157–171. https://doi.org/10.1016/j.jfoodeng.2013.04.001

Qu, J.-H., Cheng, J.-H., Sun, D.-W., Pu, H., Wang, Q.-J., Ma, J., 2015. Discrimination of shelled shrimp (Metapenaeus ensis) among fresh, frozen-thawed and cold-stored by hyperspectral imaging technique. LWT - Food Sci. Technol. 62, 202–209. https://doi.org/10.1016/j.lwt.2015.01.018

Ramos-Diaz, J.M., 2019. Application of NIR imaging to the study of expanded snacks containing amaranth, quinoa and kañiwa. Food Sci. Technol. 7.

Reibel, Y., Jung, M., Bouhifd, M., Cunin, B., Draman, C., 2003. CCD or CMOS camera noise characterisation. Eur. Phys. J. Appl. Phys. 21, 75–80. https://doi.org/10.1051/epjap:2002103

Rohman, A., Che Man, Y.B., 2011. The use of Fourier transform mid infrared (FT-MIR) spectroscopy for detection and quantification of adulteration in virgin coconut oil. Food Chem. 129, 583–588. https://doi.org/10.1016/j.foodchem.2011.04.070

Ropodi, A.I., Pavlidis, D.E., Mohareb, F., Panagou, E.Z., Nychas, G.-J.E., 2015. Multispectral image analysis approach to detect adulteration of beef and pork in raw meats. Food Res. Int. 67, 12–18. https://doi.org/10.1016/j.foodres.2014.10.032

Rupasinghe, R.A.A., Senanayake, S.G.M.P., Padmasiri, D.A., Ekanayake, M.P.B., Godaliyadda, G.M.R.I., Wijayakulasooriya, J.V., 2016. Modes of clustering for motion pattern analysis in video surveillance, in: 2016 IEEE International Conference on Information and Automation for Sustainability (ICIAfS). Presented at the 2016 IEEE International Conference on Information and Automation for Sustainability (ICIAfS), IEEE, Galle, Sri Lanka, pp. 1–6. https://doi.org/10.1109/ICIAFS.2016.7946550

Young, F.V.K., 1983. Palm Kernel and coconut oils: Analytical characteristics, process technology and uses. J. Am. Oil Chem. Soc. 60, 374–379. https://doi.org/10.1007/BF02543521